\documentclass[prl,nofootinbib,twocolumn,superscriptaddress]{revtex4}

\usepackage{psfrag}

\def\mysection#1{{\bf #1.} }
\def\mysections#1{{\bf #1.} }
\usepackage{amssymb}
\usepackage{amsmath}
\usepackage[dvips]{graphicx}
\usepackage{longtable}
\usepackage{verbatim}
\usepackage{amsfonts}

\newcommand{\be}{\begin{equation}}
\newcommand{\ee}{\end{equation}}
\newcommand{\bea}{\begin{eqnarray}}
\newcommand{\eea}{\end{eqnarray}}
\newcommand{\beq}{\begin{equation}}
\newcommand{\eeq}{\end{equation}}
\def\beqa{\begin{eqnarray}}
\def\eeqa{\end{eqnarray}}
\newcommand{\no}{\nonumber}
\def\lsim{\mathrel{\rlap{\lower4pt\hbox{\hskip1pt$\sim$}}
    \raise1pt\hbox{$<$}}}         %less than or approx. symbol
\def\gsim{\mathrel{\rlap{\lower4pt\hbox{\hskip1pt$\sim$}}
    \raise1pt\hbox{$>$}}}         %greater than or approx. symbol

\begin{document}
%\draft

%\preprint

\vspace*{-30mm}

\title{\boldmath CP Violation in $\tau^\pm\to\pi^\pm K_S\nu$ and
  $D^\pm\to\pi^\pm K_S$: \\
  The Importance of $K_S-K_L$ Interference}

\author{Yuval Grossman}\email{yg73@cornell.edu}
\affiliation{Institute for High Energy Phenomenology, Newman
  Laboratory of Elementary Particle Physics, Cornell University,
  Ithaca, NY 14853, USA}

\author{Yosef Nir}\email{yosef.nir@weizmann.ac.il}
\affiliation{Department of Particle Physics and Astrophysics,
  Weizmann Institute of Science, Rehovot 76100, Israel}

\vspace*{1cm}
%\date{\today}
%\pacs{12.10.Dm, 12.10.Kt, 98.80.Cq}

\begin{abstract}
  The $B$-factories have measured CP asymmetries in the $\tau\to\pi
  K_S\nu$ and $D\to K_S\pi$ modes. The $K_S$ state is identified by
  its decay to two pions at a time that is close to the $K_S$
  lifetime. Within the Standard Model and many of its extensions, the
  asymmetries in these modes come from CP violation in
  $K^0-\overline{K}^0$ mixing. We emphasize that the interference
  between the amplitudes of intermediate $K_S$ and $K_L$ is as
  important as the pure $K_S$ amplitude. Consequently, the measured
  asymmetries depend on the times over which the relevant decay rates
  are integrated and on features of the experiment.
\end{abstract}

\maketitle

%%%%%%%%%%%%%%%%%%%%
\mysection{Introduction}
The BaBar collaboration has recently announced a measurement of the CP
asymmetry in the $\tau\to\pi K_S\nu_\tau$ decay
\cite{Collaboration:2011mj}:
\beqa\label{eq:aqexp}
A_\tau&\equiv&\frac{\Gamma(\tau^+\to\pi^+K_S\bar\nu_\tau)
-\Gamma(\tau^-\to\pi^- K_S\nu_\tau)}
{\Gamma(\tau^+\to\pi^+K_S\bar\nu_\tau)
+\Gamma(\tau^-\to\pi^- K_S\nu_\tau)}\no\\
&=&(-4.5\pm2.4\pm1.1)\times10^{-3}.
\eeqa
(See \cite{Bischofberger:2011pw} for related measurements.)
The BaBar \cite{delAmoSanchez:2011zza,White:2011rm}, BELLE
\cite{Ko:2010ng}, CLEO \cite{:2009ak,:2007zt} and FOCUS
\cite{Link:2001zj} collaborations have measured the CP asymmetry in
the $D\to\pi K_S$ decay:
\beqa\label{eq:dcpdef}
A_D&\equiv&\frac{\Gamma(D^+\to K_S\pi^+)-\Gamma(D^-\to K_S\pi^-)}
{\Gamma(D^+\to K_S\pi^+)+\Gamma(D^-\to K_S\pi^-)}\no\\
&=&(-5.4\pm1.4)\times10^{-3},
\eeqa
where the numerical value is an average over the four measurements.

Assuming that direct CP violation in the $\tau$ or $D$ decay plays a
negligible role, as is the case in the Standard Model and many of its
extensions, then the asymmetries (\ref{eq:aqexp}) and
(\ref{eq:dcpdef}) arise from CP violation in $K^0-\overline{K}^0$
mixing~\cite{Lipkin:1999qz,Bigi:2005ts,Calderon:2007rg}. It is
important then to realize two facts:
\begin{enumerate}
\item The $\tau^+\,(\tau^-)$ decay produces initially a
  $K^0\,(\overline{K}^0)$ state, while the $D^+\,(D^-)$ decay produces
  initially a $\overline{K}^0\,(K^0$) state. (The color and doubly
  Cabibbo suppressed $D^+\to K^0\pi^+$ decay amplitude can be safely
  neglected.)
\item The intermediate $K_S$-state is not directly observed in the
  experiments. It is {\it defined} via a final $\pi^+\pi^-$ state with
  $m_{\pi\pi}\approx m_K$ and a time difference between the $\tau$ or
  $D$ decay and the $K$ decay $t\approx\tau_S$, where $\tau_S$ is the
  $K_S$ lifetime.
\end{enumerate}
Thus, in the absence of direct CP violation, the asymmetries depend on
the integrated decay times, and we have
\beqa \label{eq:taeps}
A_\tau(t_1,t_2)&=&-A_D(t_1,t_2)= A_\epsilon(t_1,t_2),\\
A_\epsilon(t_1,t_2)&=&\frac
{\int_{t_1}^{t_2}dt[\Gamma(K^0(t)\to\pi\pi)-\Gamma(\overline{K}^0(t)\to\pi\pi)]}
{\int_{t_1}^{t_2}dt[\Gamma(K^0(t)\to\pi\pi)+\Gamma(\overline{K}^0(t)\to\pi\pi)]},\no
\eeqa
where $K^0(t)\,(\overline{K}^0(t)$) is a time-evolved initially-pure
$K^0\,(\overline{K}^0$). The fact that $A_\tau(t_1,t_2)$ and
$A_D(t_1,t_2)$ are predicted to have opposite signs, while the
experimental measurements  (\ref{eq:aqexp}) and (\ref{eq:dcpdef})
carry the same sign is intriguing. The naive expectation that
$A_\tau=-A_D$ is excluded at $3.3\sigma$.

In this work, we derive an explicit expression for the
$A_\epsilon(t_1,t_2)$ asymmetry and its dependence on the
experimentally known mixing parameters $\epsilon$ and $\Delta m$. In
doing so, we correct for sign mistakes made in previous literature.
We argue that the theoretical prediction depends on $t_1$, $t_2$ and
on details of the experiment. Until these subtleties are taken into
consideration, it is difficult to asses the significance of
$A_\tau\neq-A_D$. 

\begin{figure}[t!]\begin{center}
%\psfrag{t}{\hspace*{-9pt}$\mathrel{\rlap{\lower24pt\hbox{$\displaystyle{t\over\tau_S}$}}}$}
\psfrag{t}{$\displaystyle{t\over\tau_S}$}
\psfrag{R}{\hspace*{-2pt}$R$}
 \includegraphics[width=77mm]{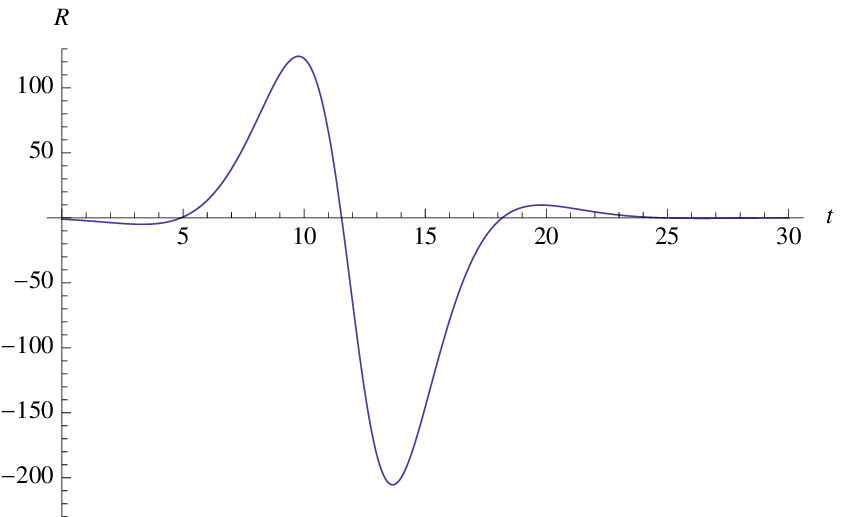}

\vspace*{7mm}
\psfrag{t}{$\displaystyle{t\over\tau_S}$}
\psfrag{R}{\hspace*{-2pt}$R$}
 \includegraphics[width=77mm]{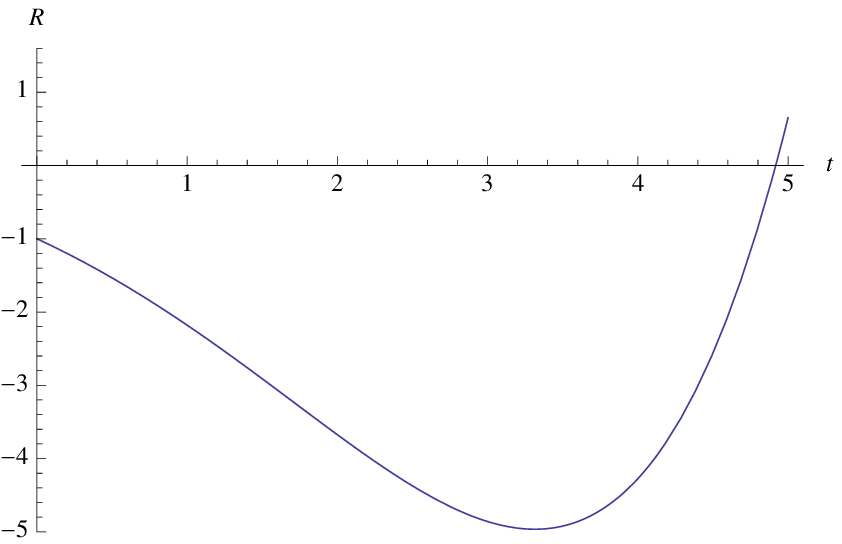}
\caption{The ratio $R$ defined in Eq. (\ref{eq:defr}) as a 
    function of time in units of $\tau_S$. In the lower plot we zoom
into short times.}
\label{fig:1a}
\end{center}
\end{figure}

%%%%%%%%%%%%%%%%%%%%
\mysection{The experimental parameters}
The two neutral $K$-meson mass eigenstates, $|K_S\rangle$ of mass $m_S$
and width $\Gamma_S$ and $|K_L\rangle$ of mass $m_L$ and width $\Gamma_L$,
are linear combinations of the interaction eigenstates $|K^0\rangle$ (with
quark content $\bar sd$) and $|\overline{K}^0\rangle$ (with quark content
$s\bar d$):
\beqa
|K_{S,L}\rangle&=&p|K^0\rangle\pm q|\overline{K}^0\rangle.
\eeqa
The average and the difference in mass and width are given by
\beqa
m\equiv\frac{m_S+m_L}{2},&\qquad
&\Gamma\equiv\frac{\Gamma_S+\Gamma_L}{2},\no\\[3pt]
\Delta m\equiv m_L-m_S,&\qquad&\Delta\Gamma\equiv\Gamma_L-\Gamma_S,\no\\[5pt]
x\equiv\frac{\Delta m}{\Gamma},&\qquad &y\equiv\frac{\Delta\Gamma}{2\Gamma}.
\eeqa
The decay amplitudes into a final state $\pi\pi$ are defined as
\beq
A_{S,L}\equiv\langle \pi\pi|{\cal H}|K_{S,L}\rangle.
\eeq
The relevant CP violating parameters are defined as
\beq
\frac{|p|^2-|q|^2}{|p|^2+|q|^2}\approx2{\cal R}e(\epsilon),\qquad
\frac{A_L}{A_S}\approx\epsilon,
\eeq
where in the first approximation we neglected a correction of relative
order $|\epsilon|^2$, and in the second a correction of relative order
$\epsilon^\prime/\epsilon$.
We obtain:
\beqa\label{timdepr}
\Gamma_{\pi\pi}(t)&\equiv&\Gamma(K^0(t)\to\pi\pi)\\
&=&\frac{|A_S|^2}{4|p|^2}\left(e^{-\Gamma_St}+
  |\epsilon|^2e^{-\Gamma_Lt}+2{\cal R}e(
  e^{i\Delta mt-\Gamma t}\epsilon^*)\right)\no\\[2pt]
\bar\Gamma_{\pi\pi}(t)&\equiv&\Gamma(\overline{K}^0(t)\to\pi\pi)\no\\
&=&\frac{|A_S|^2}{4|q|^2}\left(e^{-\Gamma_St}+
  |\epsilon|^2e^{-\Gamma_Lt}-2{\cal R}e(
  e^{i\Delta mt-\Gamma t}\epsilon^*)\right).\no
\eeqa
For the difference and the sum of these rates, we obtain
\beqa\label{eq:dift}
\frac{D_{\pi\pi}(t)}{N}&\equiv&\frac{\Gamma_{\pi\pi}(t)-\bar\Gamma_{\pi\pi}(t)}{N} \\[2pt]
&=&-2{\cal R}e(\epsilon)\left(e^{-\Gamma_St}+
  |\epsilon|^2e^{-\Gamma_Lt}\right)\no\\[2pt]
&&+2e^{-\Gamma t}\left({\cal
    R}e(\epsilon)\cos(\Delta mt)+{\cal I}m(\epsilon)\sin(\Delta
  mt)\right),\no\\[6pt]
%\eeqa
%while for the sum we obtain
%\beqa
\label{eq:sum}
\frac{S_{\pi\pi}(t)}{N}&\equiv&
\frac{\Gamma_{\pi\pi}(t)+\bar\Gamma_{\pi\pi}(t)}{N}\\[2pt]
&=&e^{-\Gamma_St}+
  |\epsilon|^2e^{-\Gamma_Lt}
-4{\cal R}e(\epsilon)e^{-\Gamma t}\no\\[2pt]
 &&\times\left({\cal
    R}e(\epsilon)\cos(\Delta mt)+{\cal I}m(\epsilon)\sin(\Delta
  mt)\right),\no
\eeqa
where 
\beq
N=|A_S|^2 \frac{|p|^2+|q|^2}{4|p|^2|q|^2}. 
\eeq 
For the sum $S_{\pi\pi}(t)$ of Eq. (\ref{eq:sum}), the interference
(and the pure $K_L$) terms are suppressed by ${\cal O}(\epsilon^2)$
compared to the pure $K_S$ term. For the difference $D_{\pi\pi}(t)$ of
Eq. (\ref{eq:dift}), however, this is not the case. The ratio between
the second (interference) and first (non-interference) terms in
$D_{\pi\pi}(t)$, 
\beq\label{eq:defr}
R(t)\equiv-\frac{e^{-\Gamma t}\left(\cos(\Delta mt)+\frac{{\cal
        I}m(\epsilon)}{{\cal R}e(\epsilon)}\sin(\Delta
  mt)\right)}{e^{-\Gamma_St}+ |\epsilon|^2e^{-\Gamma_Lt}},
\eeq
is plotted in Fig.~\ref{fig:1a} as a function of time. In the figure
we can observe the following features:
\begin{enumerate}
\item Even at very early times, the interference term is not
  negligible compared to the pure $K_S$ term. For example, at $t=0$,
  $R= -1$.    
\item In the approximations [good to ${\cal O}(5\%)$] that
  ${\cal R}e(\epsilon)\approx {\cal I}m(\epsilon)$ and $x\approx-y$,
  the ratio changes sign when
  $\tan[\frac12(t/\tau_S)]=-1$, namely $t/\tau_S=3\pi/2+2n\pi\
  (n=0,1,2,\ldots)$.  
\item For times early enough that the pure $K_L$ term can be
  neglected $(t\ll12\tau_S$), $R$ reaches a minimum at
  $t/\tau_S\sim\pi$, $R\sim-e^{\pi/2}$, and a maximum at
  $t/\tau_S\sim3\pi$, $R\sim+e^{3\pi/2}$. 
\end{enumerate}

Since the CP asymmetry depends on the time at which the kaon decays,
the final measurement is sensitive to the experimental cuts. To
incorporate these cuts, we need to take into account not only the
efficiency as a function of the kaon decay time, but also the kaon
energy in the lab frame to account for time dilation. We parametrize
all of these experiment-dependent effects by a function $F(t)$ such
that $t$ is the time in the kaon rest frame and $0 \le F(t) \le 1$. We
emphasize that this function must be determined as part of the
experimental analysis. The experimentally measured asymmetry is thus
given by the convolution of the bare asymmetry with $F$:
\beq\label{eq:aexp}
A_\epsilon = \frac{\int_{0}^{\infty} F(t)
  D_{\pi\pi}(t)\,dt}{\int_{0}^{\infty} F(t) S_{\pi\pi}(t)\,dt}\,.
\eeq

\begin{figure}[!t]
\psfrag{t}{$\displaystyle{t_1\over\tau_S}$}
\psfrag{A}{\hspace*{-2pt}$A_\epsilon$}
 \includegraphics[width=77mm]{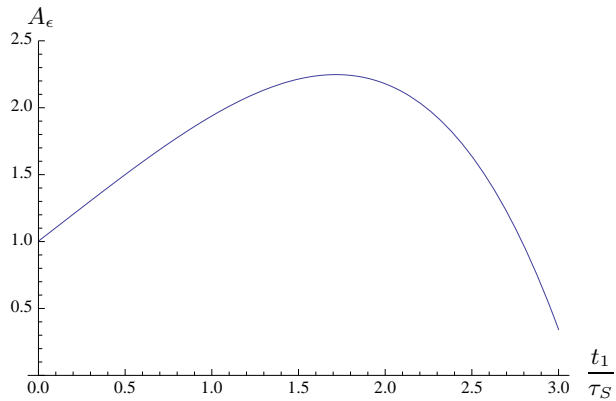}
\caption{$A_\epsilon(t_1,t_2)$ given in Eq. (\ref{eq:aeps}) in units
  of $[2{\cal R}e(\epsilon)]$ as a function of $t_1/\tau_S$ for
  $t_2=10\tau_S$.}
\label{fig:2a}
\end{figure}

\begin{figure}[!t]
\psfrag{t}{$\displaystyle{t_2\over\tau_S}$}
\psfrag{A}{\hspace*{-2pt}$A_\epsilon$}
 \includegraphics[width=77mm]{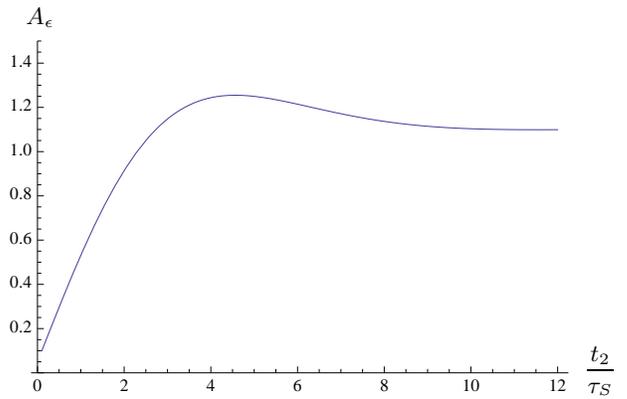}
\caption{$A_\epsilon(t_1,t_2)$ given in Eq. (\ref{eq:aeps}) in units
  of $[2{\cal R}e(\epsilon)]$ as a function of $t_2/\tau_S$ for
  $t_1=0.1\tau_S$.}
\label{fig:3a}
\end{figure}

While we do not have the function $F(t)$, it is reasonable to
approximate it by a double step function,
\beq
F(t)=\begin{cases}1 &t_1<t<t_2 \\
  0 & \mbox{otherwise}.\end{cases}
\eeq
In this case the experimentally measured asymmetry, $A_\epsilon$
defined in Eq.~(\ref{eq:aexp}), coincides with the theoretical one,
$A_\epsilon(t_1,t_2)$ defined in Eq.~(\ref{eq:taeps}). When
$t_2\ll\tau_L$, we can safely neglect terms of ${\cal O}(\epsilon^2)$:
\beqa
&&A_\epsilon(t_1,t_2)=-2{\cal R}e(\epsilon)\\
&&\times\left[1-\frac
{\int_{t_1}^{t_2}dt e^{-\Gamma t}\left(\cos(\Delta m t)
    +\frac{{\cal I}m(\epsilon)}{{\cal R}e(\epsilon)}\sin(\Delta m
    t)\right)}
{\int_{t_1}^{t_2}dt e^{-\Gamma_S t}}\right].\no
\eeqa
Neglecting direct CP violation, we can use the model independent
relation \cite{Grossman:2009mn} 
\beq
\frac{{\cal I}m(\epsilon)}{{\cal R}e(\epsilon)}=-\frac{x}{y},
\eeq
to obtain
\beqa\label{eq:aeps}
&&\!\!\!\!\!\!\!A_\epsilon(t_1,t_2)=-2{\cal R}e(\epsilon)\huge[1\\
&&-\frac{2(1-x^2/y)}{1+x^2}
\frac{e^{-\Gamma t_1}\cos(\Delta m t_1)
  -e^{-\Gamma t_2}\cos(\Delta m t_2)}{e^{-\Gamma_S
  t_1}-e^{-\Gamma_S t_2}}\no\\
&&+\left.\frac{2(x+x/y)}{1+x^2}
\frac{e^{-\Gamma t_1}\sin(\Delta m t_1)
  -e^{-\Gamma t_2}\sin(\Delta m t_2)}{e^{-\Gamma_S
  t_1}-e^{-\Gamma_S t_2}}\right].\no
\eeqa
A particularly simple result arises when $t_1\ll\tau_S$ and
$\tau_S\ll t_2\ll\tau_L$, so that we can take $e^{-\Gamma_S t_1}=1$,
$e^{-\Gamma_S t_2}=0$, and $\cos(\Delta m t_1)=1$. In addition we use
$y\simeq-1$, and obtain 
\beq
A_\epsilon(t_1\ll\tau_S,\tau_S\ll t_2\ll\tau_L)=+2{\cal
R}e(\epsilon)\approx +3.3\times 10^{-3},
\eeq
where in the last step we used the experimental value.
In Figs.~\ref{fig:2a} and \ref{fig:3a} we investigate the dependence
of $A_\epsilon(t_1,t_2)$ on the choice of $t_1$ and $t_2$. In
Fig.~\ref{fig:2a} we plot $A_\epsilon(t_1,t_2)/(2{\cal R}e(\epsilon))$
as a function of $t_2$ for $t_1=\tau_S/10$. In Fig.~\ref{fig:3a} we
plot $A_\epsilon(t_1,t_2)/(2{\cal R}e(\epsilon))$ as a function of
$t_1$ for $t_2=10\tau_S$.
%
%%%%%%%%%%%%%%%%%%%%%%%%%%%%%%%%%%%%%%%%%%%%%%%%%%%%%%%
%%%%%%%%%%%%%%%%%%%%%%%%%%%%%%%%%%%%%%%%%%%%%%%%%%%%%%%
We emphasize the following points:
\begin{enumerate}
\item For $t_2$ large enough that the $e^{-\Gamma t_2}$ term is
  negligible, and for $t_1/\tau_S\ll1$, we have
\beqa\label{eq:aepsone}
A_\epsilon(t_1,t_2)\approx+2{\cal
R}e(\epsilon)(1+t_1/\tau_S). 
\eeqa
This linear rise with $t_1$, which can be clearly seen in
Fig. \ref{fig:2a}, is a result of ``losing'' a fraction
$t_1/\tau_S$ of the time independent pure $K_S$ term in the asymmetry.
\item For $t_1$ fixed and small, $A_\epsilon$ reaches a maximum at
  around $t_2=(3\pi/2)\tau_S$, and then, for higher $t_2$, converges to
  its asymptotic value of Eq. (\ref{eq:aepsone}). These features can
  be clearly seen in Fig. \ref{fig:3a}. The maximum is
  enhanced by a factor of about $(1+\sqrt{2}\exp(-3\pi/4))\approx1.13$
  compared to the asymptotic value. 
\end{enumerate}

Let us comment on previous relevant literature. The idea to measure
the CP asymmetry of Eq. (\ref{eq:aqexp}) was first made in Ref.
\cite{Bigi:2005ts}. In this beautiful work, the importance of the
interference term in restoring the CPT constraint is explained.
Indeed, the BaBar paper \cite{Collaboration:2011mj} compare their
measurement to the prediction given in Eq. (7) of Ref.
\cite{Bigi:2005ts}. We note, however, that both Eq. (6) and Eq. (7) of
Ref. \cite{Bigi:2005ts} have a sign mistake: Both the ``pure $K_L$''
term and the ``pure $K_S$'' terms give $|q|^2-|p|^2\simeq-2{\cal
  R}e(\epsilon)$. Yet, when the interference term is taken into
account, it approximately reverses the sign of the pure $K_S$ result.
Correcting the sign of Eq. (7) in \cite{Bigi:2005ts} and taking into
account the interference term combine to approximately reproduce the
numerical prediction quoted in this equation. Further analysis of this
asymmetry is given in Ref. \cite{Calderon:2007rg}. Here the fact that
the interference term practically reverses the sign of the ``pure
$K_S$'' asymmetry is nicely pointed out, yet several sign mistakes lead
to a wrong sign in the final prediction, see their Eq. (14).
The idea to measure the CP asymmetry of Eq. (\ref{eq:dcpdef}) was first
made in Ref. \cite{Lipkin:1999qz}. The interference term is not
discussed in this work.

%%%%%%%%%%%%%%%

%%%%%%%%%%%%%%%%%%%%
\mysection{Conclusions}
CP asymmetries of ${\cal O}(10^{-3})$ in the $\tau^\pm\to\pi^\pm
K_S\nu$ and $D^\pm\to\pi^\pm K_S$ decays are predicted within the
Standard Model as a result of CP violation in $K^0-\overline{K}^0$
mixing. A violation of the SM predictions would imply direct CP
violation in $\tau$ and/or $D$ decays.

The kaon is identified via final two pions with invariant mass
$m_{\pi\pi}\sim m_K$ and decay time $t\sim\tau_S$. In the total decay
rate, the contribution of intermediate $K_S$ is strongly dominant. In
the CP asymmetry, however, the $K_L-K_S$ interference term is as
important as the pure $K_S$ term.

As a consequence of this situation, the asymmetry depends sensitively
on the decay time interval over which it is measured, and on details
of the experiment. The exact SM prediction can be obtained only if the
relevant experimental features are taken into consideration.
Generically, we expect the measured asymmetry to be opposite in sign
and larger in magnitude than the asymmetry that would arise from the
pure $K_S$ contribution.

While we focused here on only the two specific examples of $D$ and
$\tau$ decays, the analysis above applies to any measurement of a CP
asymmetry that involves $K_S$ in the final state. In particular,
similar effects should eventually be taken into account in the
determination of the angle $\gamma$ of the unitarity triangle based on
$B\to D K$ decays, if the kaon is identified as a
$K_S$~\cite{Gronau:2004gt} or if the $D$ decays into a final state
with a $K_S$ such as with the Dalitz decay $D \to \pi^+\pi^-
K_S$~\cite{Giri:2003ty}. Another case where the effect should
eventually be included is in the determination of $D -\overline{D}$
mixing using $D$ decays into $K_S$. In this case our formalism cannot
be directly applied, because at $t=0$ the kaon state is not a pure
$K^0$ (or $\overline{K}{}^0$), and adjustment to such cases is needed.

The measured asymmetry in $D$ decay seems very consistent with the SM
prediction, while the measured asymmetry in $\tau$ decay seems
different from the SM prediction by at least $3\sigma$. In view of the
potential implications for new, CP violating physics, we urge the
experimenters to take into account the subtleties that we point out,
and provide not only the measured value of the asymmetry, but also the
theoretical prediction which depends on specific experimental features. 

%%%%%%%%%%%%
\mysections{Acknowledgments} 
We thank Abner Soffer for discussions. This work is supported by the
United States-Israel Binational Science Foundation (BSF), Jerusalem,
Israel.  The work of YG is supported by the NSF grant PHY-0757868. The
work of YN is supported by the Israel Science Foundation founded by
the Israel Academy of Sciences and Humanities, and by the
German-Israeli foundation for scientific research and development
(GIF).

%\vspace*{-5mm}
%%%%%%%%%%%%%%%%%%%%%

\end{document}